\documentclass[aps,prd,tightenlines,showpacs,amsmath,superscriptaddress]{revtex4} 
\usepackage{graphics}

\begin{document}

\author{M. Shaposhnikov}
\affiliation{Institut de Th\'eorie des Ph\'enom\`enes Physiques, Ecole Polytechnique F\'ed\'erale de Lausanne, CH-1015 Lausanne,  Switzerland}
\author{P. Tinyakov}
\affiliation{Service de Physique Th\'eorique, Universit\'e Libre de Bruxelles, CP225, bld.~du Triomphe, B-1050 Bruxelles, Belgium} 
\author{K. Zuleta}
\affiliation{Institut de Th\'eorie des Ph\'enom\`enes Physiques, Ecole Polytechnique F\'ed\'erale de Lausanne, CH-1015 Lausanne,  Switzerland}

\title{Quasilocalized gravity without asymptotic flatness}

\date{\today}

\begin{abstract}%
We present a toy model of a generic five-dimensional warped
 geometry in
which the 4D graviton is not fully localized on the brane. 
Studying the tensor sector of metric perturbation around this background, we find that its contribution to
the effective gravitational potential is of 4D type ($1/r$) at the intermediate scales and 
that at the large scales it  becomes  $1/r^{1+\alpha}$, $0<\alpha\leq 1$ being a function of the parameters of the model
($\alpha=1$ corresponds to the asymptotically flat geometry). Large-distance behavior of the potential is 
therefore not necessarily five-dimensional. Our
analysis applies also to the case of quasilocalized massless particles
other than graviton.
\end{abstract}


\maketitle

\section{Introduction}

The idea that we could live on a brane embedded in a higher-dimensional
space dates back to~Refs.~\cite{shap,akama} where it was proposed  that the universe 
be modeled as a topological defect (a domain wall or a  string) in a higher-dimensional space-time. The key
feature of these models is the localization of matter on the brane, which
makes the theory look four-dimensional at low energies. In the setup of Ref.~\cite{shap}
this was achieved by means of Yukawa interaction between the domain wall and the
fermion field. The latter has a zero mode localized on the brane, as well
as the continuum modes separated from the zero mode by a gap.  At low
energies, only the zero mode remains in the spectrum and plays the role of a
four-dimensional particle. It was suggested later~\cite{visser,squires} that the gravitational
attraction may also trap particles on the brane.

This idea has been the subject of renewed interest since it has been
shown~\cite{rs} that the graviton itself can be trapped on the brane by the
gravitational force, leading to effectively four-dimensional gravity.
Considering a brane embedded in a 5D anti-de Sitter (AdS$_5$) space-time,
the authors of~Ref.~\cite{rs} have found that in this ``warped'' geometry
the 5D graviton has a zero-energy bound state, which, being localized on
the brane, can be interpreted as the ordinary 4D graviton. Even though
there is no gap separating the zero mode of graviton from the continuum,
the law of gravity on the brane is essentially Newtonian, the
corrections yielded by light continuum modes modifying this behavior only at
very short distances.

Soon after, a model was proposed~\cite{grs} in which 4D Newton's law is reproduced
only at intermediate scales and gets modified at very large scales.  
The geometry considered consisted of a flat 5D space glued onto a
portion of~AdS$_5$ space.  It has been shown that, even though there is no
localized graviton, Newton's gravity  can still be recovered on the brane at intermediate
scales.  At short distances, the gravitational potential receives
corrections similar to those encountered in the model of~Ref.~\cite{rs}.  At
very large scales, the extra dimensions ``open up'' and the gravitational potential receives ``five-dimensional'' contributions from the tensor modes, that is contributions  scaling as~$r^{-2}$,  $r$ being the distance between sources.
This behavior was related~\cite{ceh,dgp-comment_1} to the
presence of a resonant mode -- a ``quasi-bound state'' -- at zero energy,
with lifetime proportional to the long-distance scale of the model (see
also~Ref.~\cite{grs} itself).  Subsequent works~\cite{dgp-comment_1,dgp-comment_2,grs_comment,prz} made  manifest  the importance of the scalar sector (and namely the radion mode) in the description of the  effective gravity on the brane in models with metastable gravitons. Indeed, being constructed from the massive modes, the propagator of the effective 4D graviton  is likely  to have an incorrect tensor structure, 
 leading to predictions inconsistent with the observations~\cite{dgp-comment_1,prz}. Coupling to the trace of the energy-momentum tensor, radion  contributes to the graviton propagator and thus makes it possible to recover  Einstein  gravity at the intermediate distances~\cite{grs_comment,prz}. Model presented in~\cite{grs} has however a serious drawback: it  requires the use of the negative tension branes -- or, in other words, violation of the condition of positivity of energy~\cite{witten}. It was shown~\cite{prz} that this forces radion to  become a ghost~\cite{dgp-comment_2,prz} and causes the effective gravity at very large distances to become scalar anti-gravity~\cite{grs_comment}. While this behavior occurs at the experimentally inaccessible scales, presence of  negative norm states certainly raises doubts concerning the consistency of the model.

An alternative brane model producing an infrared modification of gravity 
 was constructed in~\cite{dgp_1} using a thin brane embedded in five-dimensional Minkowski space. The action included a four-dimensional Einstein term on the brane, induced by radiative
corrections due to the matter fields localized there. This term was shown
to dominate the gravitational interactions of the brane matter at
intermediate scales, giving  four-dimensional gravity at intermediate scales, while 
 at  large scales it became five-dimensional.  
 This  setup was subsequently
extended to higher dimensions~\cite{dgp_2} and its generalizations to
smooth branes were also considered~\cite{dgp_follows_2,kolanovic}.
Recently, the idea of induced gravity was applied to the warped space-times
\cite{porrati-strong_coupling,dubovsky,padilla}. 
As yet, the setups with induced gravity 
are not entirely satisfactory: most of 
 the models suffer from ghosts and/or
strong gravity
problems~\cite{porrati-strong_coupling,rub-strong_coupling,dub-rubakov}.
Nevertheless, the idea of infrared modification of gravity remains very
appealing as it is thought~\cite{witten,dgp_follows_3} that it could be a solution
to the cosmological constant problem.

The feature shared by these models 
is that at ultra-large scales gravity becomes higher-dimensional, that is the gravitational potential scales as
$r^{-(1+N)}$, $r$ being the distance between sources and $N$ the number of
extra dimensions (at least as far as the tensor sector is concerned).  It is natural to wonder whether this feature is inherent
to the quasilocalized gravity on branes embedded in
higher dimension space-times or if it could exhibit some different, more
general, behaviors. More generally, one may ask what  kinds of  large-distance
 modifications of gravity are in principle possible in this setup and under
what conditions they can be realized.

In order to cast some light on this issue, we study in this paper tensor perturbations of the metric around a fairly generic five-dimensional warped  background, in which the graviton is not fully localized on the brane (becoming a metastable state). To some extent,  our model can be considered as a toy model of quasilocalized gravity. Of course, as shows the example of~\cite{grs}, studying only the tensor sector of metric perturbations is unlikely to give a full description  of effective gravity on branes and we do not aspire to present a realistic solution for quasilocalization of gravity. The aim of this work is rather to signal a peculiar behavior of the potential yielded by the tensor perturbations in such kinds of backgrounds. For the sake of simplicity, throughout the paper  we will loosely refer to the  contribution of the tensor modes to the gravitational potential as to the potential itself, bearing in mind that in any concrete model one should account for the contribution from the scalar sector as well. Same applies to our description of  gravitational waves.

We find that  our model gives  4D gravitational potential  at intermediate scales -- quite similarly to the situation encountered in the model of~Ref.~\cite{grs}. This ``quasilocalization'' of gravity is a generic feature of the geometry which we consider -- whether or not the
embedding space-time is asymptotically flat. At large distances, we observe a novel behavior: the 
potential gets modified and is of type~$1/r^{1+\alpha}$, $0<\alpha\leq 1$ being a function of the
 parameters of the model. It becomes~5D ($\alpha=1$) only when the geometry is
asymptotically flat. Thus, the gravitational potential 
does not necessarily
exhibit~5D behavior at very large distances. 

Although throughout this paper we use the example and the language of gravity, our analysis is in fact quite universal, in that it is not constrained to the gravity sector, but can apply  to the matter sector as well. Indeed, our considerations are based on studying the properties of a  Shr\"odinger-like equation governing the behavior of the metric perturbations and, by consequence, the effective physics on the brane. Actually, irrespective of the spin of the matter fields present in the bulk, the mass spectrum of the  effective four-dimensional theory is determined by a Schr\"odinger-like equation of some type (or a system of thereof). Using this fact, several authors have studied the  localization of different matter fields in the AdS$_5$ background of Ref.~\cite{rs} and its variations: It was shown that, similarly to gravitons, the spectrum of  massless bulk scalars~\cite{bajc,giddings} consists of  a  localized zero mode, followed by the continuum of arbitrarily light states. Massless fermions~\cite{bajc} can also be localized on the brane and  mechanisms to localize gauge fields were proposed \cite{drt_2,oda,akhmedov}. Localization of massive scalars  and fermions was discussed in Ref.~\cite{drt}  (see also Ref.~\cite{ringeval}), where it was shown that while there are no truly localized states (massive or massless), the setup allows for metastable massive particles on the brane, having a small, but finite, probability of tunneling into the bulk. In the scalar case, these resonant states were shown to give  $1/r$ contribution to the effective static potential on the brane, while the light states of the continuum give a power-law behavior at large distances. Our analysis completes the picture, revealing that, under certain conditions, in the absence of true bound states and massive resonances, it is possible to have an ``almost localized'' massless state, inducing a $1/r$ potential on the brane evolving towards a (fractional) power-law behavior at large distances.         

\section{General Setup}\label{setup}

Let us begin with the general setup.  We consider 5D ``warped'' space-times
preserving~4D Poincar\'e symmetry:
\begin{equation}
ds^2=e^{-A(z)}\left(\eta_{\alpha\beta} dx^\alpha dx^\beta-dz^2\right) \ ,
\label{metric}
\end{equation} 
where the ``warp factor''~$A(z)$ in an even and non-decreasing function of~$z$.

We suppose that the ordinary matter is localized on the brane centered at
$z=0$. In order to study the nature of the gravitational interactions
experienced by this matter, we need to study perturbations around the
background metric. We will consider exclusively the tensor perturbation~$h_{\alpha\beta}(x,z)=g(z)^{3/2}\tilde{h}_{\alpha\beta}(x)u(z)$, $\alpha$ and~$\beta$ being the four-dimensional tensor indices and~$g(z)$ metric determinant.
The behavior of this perturbation  
 is governed by the Schr\"odinger-like equation
\begin{equation}  
\label{schro}
-\frac{d^2u_\mu(z)}{dz^2}+V(z)u_\mu(z)=\mu^2u_\mu(z)\ ,
\end{equation}
where~$-\partial^\rho\partial_\rho \tilde{h}_{\alpha\beta}(x)=\mu^2 \tilde{h}_{\alpha\beta}(x)$ and the potential is given in terms of the warp factor:
\begin{equation}
\label{pot} 
V(z)=\frac{9}{16}A^\prime(z)^2-\frac{3}{4}A^{\prime\prime}(z) \ ,
\end{equation}
which is a form familiar from the supersymmetric quantum mechanics with the
``superpotential'' $ W(z)=\frac{3}{4}A^{\prime}(z)\ .  $ The factorization
of the Hamiltonian:
\begin{equation}
-\frac{d^2}{dz^2}+V(z)=\left(-\frac{d}{dz}+W(z)\right)\left(\frac{d}{dz}+W(z)\right)
\end{equation}
ensures that the zero-energy mode
\begin{equation}
u_0(z)=\exp\left[-\frac{3}{4}A(z)\right]
\end{equation}
is the ground state of the system.

In this way, studying 
the influence of the tensor modes on the effective gravity on the brane
is reduced to studying the properties of the spectrum of the one-dimensional
quantum mechanical system~$(\ref{schro})$.  The quantity of particular
interest is the spectral density~$\rho(\mu)=|u_\mu(0)|^2$, as it enters the
retarded Green's function~$G_R(x,z=0;x^\prime,z^\prime=0)$ which determines
the gravitational interactions on the brane and consequently the observable
physical quantities, such as the induced gravitational potential and
radiation of gravitational waves.  Let us remind that the
(five-dimensional) retarded Green's
function~$G_R(x,z=0;x^\prime,z^\prime=0)$ of the tensor sector of the linearized Einstein
equations can be constructed from the full set of eigenmodes of
Eqn.~$(\ref{schro})$ via spectral decomposition:
\begin{widetext}
\begin{equation}
\label{green_f}
G_R(x,z;x^\prime,z^\prime)=\int_0^\infty d\mu \ u_{\mu}(z)
u_{\mu}(z^\prime)\int \frac{d^4
p}{(2\pi)^4}\frac{e^{-ip\cdot(x-x^\prime)}}{\mu^2-p^2-i\varepsilon p^0} \ .
\end{equation}
\end{widetext}

\section{Model}

The purpose of this work is not to present any particular realistic setup
for the quasilocalized gravity, but rather to study general properties of
warped space-times in which this phenomenon can arise and determine what are the possible 
large distance behaviors of  the gravitational potential (induced by the tensor modes).   
With this aim in mind, we construct and study a toy model for a generic geometry allowing
(quasi)localization of gravity.  Concretely, we consider a class of
potentials $V(z)$ with ``volcano'' shape, possibly with tails -- for which
we choose~$1/z^2$ fall-off, familiar from the Randall-Sundrum
model~\cite{rs}. Knowing that the potential in question necessarily follows from a
``superpotential'', we start our study by defining one of the following
form:
\begin{equation}
\label{w}
W(z)=-W(-z)=\begin{cases}
a\tan(az)\displaystyle{\frac{}{}} & z<z_0\ , \\
\displaystyle{\frac{}{}}b \tanh\left[b\left(z_2-z\right)\right] & z_0\leq z \leq z_c\ ,\\
\displaystyle{\frac{c}{z-z_1}} & z\geq z_c\ .\\
\end{cases}
\end{equation}
The motivation for choosing this particular form of superpotential is  twofold: not only does it yield a 
simple potential for which an exact solution of the Schr\"odinger equation can be determined, but also allows 
 a unified description of the (quasi)localization, naturally including regimes already discussed in the literature~\cite{rs,ceh}. 
Our choice of superpotential corresponds in general to geometries which are not
asymptotically flat -- except when the limit~\mbox{$c=0$} is considered.
To ensure that the resulting potential does not have singularities stronger
than finite jumps we demand that~$W(z)$ be continuous. This requirement
entails relations between the constants:
\begin{equation}
\label{cont_w_1}
\begin{cases}
\displaystyle{b\tanh\left[b\left(z_2-z_0\right)\right]=a\tan(az_0)} \ , \\
\displaystyle{b\tanh\left[b\left(z_2-z_c\right)\right]=\frac{c}{z_c-z_1}} \ .  
\end{cases}
\end{equation}
The potential generated by the superpotential~$(\ref{w})$ is symmetric and
has the desired ``volcano-like'' shape. It is given by:
\begin{equation}
V(z)=-W^\prime(z)+W(z)^2=
\begin{cases}
-a^2 \displaystyle{\frac{}{}} & |z|<z_0\ , \\
\displaystyle{\frac{}{}}b^2  &   z_0<|z|<z_c\ , \\
\displaystyle{\frac{c(1+c)}{(|z|-z_1)^2}} & |z|>z_c\ .\\
\end{cases}
\end{equation}
\begin{figure*}
\includegraphics{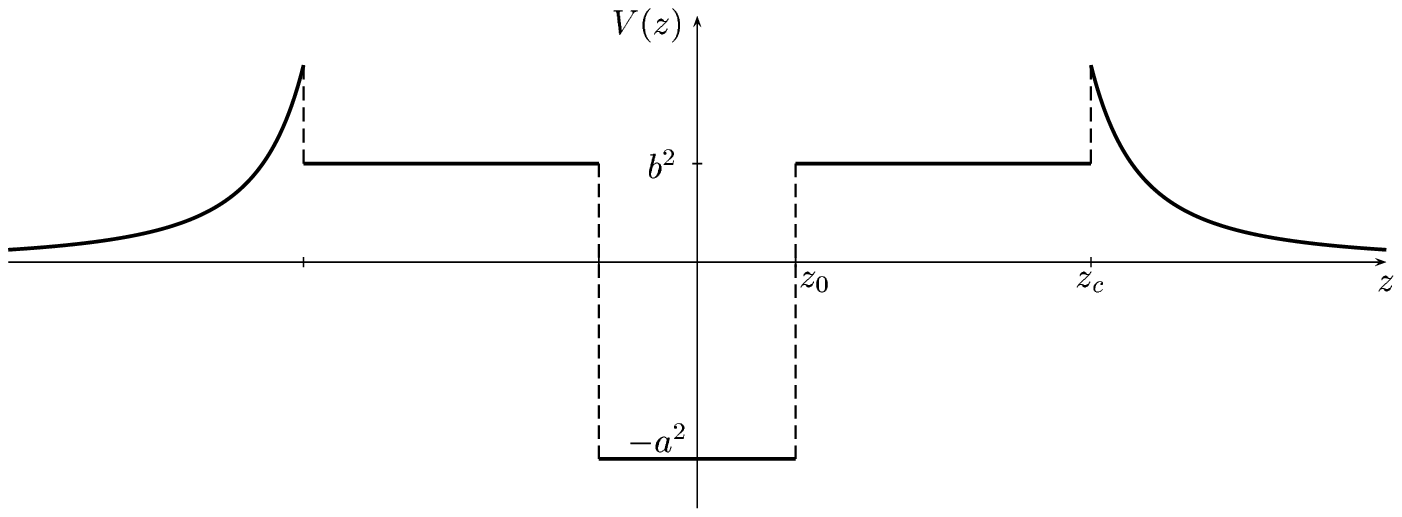}
\caption{\small Potential $V(z)$.}
\label{fig:pot}
\end{figure*}
\noindent The  parameters  of the potential must satisfy the condition following from the constraints~$(\ref{cont_w_1})$ by eliminating~$z_2$:
\begin{equation}
\label{cont_w_2}
c\left[1-\frac{a}{b}\tan(az_0)\tanh\left[b\left(z_c-z_0\right)\right]\right]=\left(z_c-z_1\right)\left[a\tan(az_0)-b\tanh\left[b\left(z_c-z_0\right)\right]\right] \ . 
\end{equation} 
In order for~$(\ref{cont_w_1})$ to remain valid we must also have 
$$
b\geq a\tan(az_0)\qquad \qquad \mbox{and}\qquad \qquad b\geq\frac{c}{z_c-z_1} \ .
$$
In spite of its simplicity, the potential~$V(z)$ presents a range of behaviors, depending on the choice of parameters, and  thus  generates  different localization schemes for gravity.    
To start, in the limit~$z_c\to \infty $, $V(z)$ becomes a simple rectangular potential well, allowing a localized symmetric zero-mode of the form:
\begin{equation}
u^\infty_{0}(z) ={\cal{C}}^\infty_0 
\begin{cases}
 \displaystyle{\cos(az)\frac{}{}} & z<z_0 \\
 \displaystyle{\cos(az_0)e ^{-b (z-z_0)}\frac{}{}} & z\geq z_0 
\end{cases}
\end{equation}
where
\begin{equation}
\label{C_infty}
\left({\cal{C}}^\infty_0\right)^{-2} =z_0 +\frac{1}{a}\cot(az_0)\ ,
\end{equation}
followed by the continuum spectrum starting at~$\mu=b$.

For finite $z_c$,
 the asymptotics of the potential makes the continuum descend towards~$\mu=0$. Condition~$(\ref{cont_w_2})$ guarantees the absence of  growing contribution to the~$\mu=0$ wave function in the region~$z>z_c$, where it consequently falls off  as~$(z-z_1)^{-c}$. 
When  $c> 1/2$, it  is therefore still normalizable and, at the bottom of the continuum, we still have a localized zero-mode. The  resulting gravity on the brane will then be of the type described in~Ref.~\cite{rs}. 
When $c\leq 1/2$, $u_{0}(z)$ is no longer  normalizable and the corresponding state can be qualified as ``quasi-bound''. 
In the particular case~$\nu=1/2$, $V(z)$ is identical to the ``volcano box'' potential used in~\cite{ceh} in order  to explain the quasilocalization of gravity in the asymptotically flat model~\cite{grs}. We claim  that quasilocalization of gravity  is present for the whole family of potentials with~$c\leq 1/2$. Let us from now on concentrate on this family. 

\section{Wave-function and spectral density}

\noindent The  solution for a symmetric continuum wave function has the form:
\begin{widetext}
\begin{equation}
\label{wave_f}
u_{\mu}(z) = u_{\mu}(0)
\begin{cases}
\displaystyle{\cos(kz)\frac{}{}} & z\leq z_0\ , \\
\displaystyle{D_1(\mu) e ^{-\kappa z}+ D_2(\mu)e^{\kappa z}\frac{}{}} &  z_0<z<z_c\ , \\
\displaystyle{\sqrt{z-z_1}\left[\frac{}{}B_1(\mu)J_\nu\left(\mu(z-z_1)\right)+B_2(\mu)N_\nu\left(\mu(z-z_1)\right)\right]} & z\geq z_c\ .
\end{cases}
\end{equation}
where 
\begin{equation}
k=\sqrt{\mu^2+a^2}\ ,\quad \kappa=\sqrt{b^2-\mu^2}\quad \mbox{and} \quad \nu=c+\frac{1}{2}\ .
\end{equation}
\end{widetext}
The coefficients in~$(\ref{wave_f})$ are determined, as usual,
by the requirement of continuity of the wave function and its
 derivative at~$z=z_0$ and~$z=z_c$: 
\begin{widetext}
\begin{subequations}
\begin{align}
B_1(\mu)&=\frac{\pi}{2}\sqrt{z_c-z_1}\left\{\left[\kappa f_{-}(\mu)+\frac{2\nu+1}{2(z_c-z_1)}f_{+}(\mu)\right]N_\nu(\mu(z_c-z_1))-\mu f_{+}(\mu)\, N_{\nu+1}(\mu(z_c-z_1))\right\} \ , \\
B_2(\mu)&=-\frac{\pi}{2}\sqrt{z_c-z_1}\left\{\left[\kappa f_{-}(\mu)+\frac{2\nu+1}{2(z_c-z_1)}f_{+}(\mu)\right]J_\nu(\mu(z_c-z_1))-\mu f_{+}(\mu)\, J_{\nu+1}(\mu(z_c-z_1))\right\}\ ,\\
f_{+}(\mu)&= D_1(\mu)e^{-\kappa z_c} + D_2(\mu)e^{\kappa z_c}=\cos(kz_0)\cosh\left[\kappa(z_c-z_0)\right]-\frac{k}{\kappa}\sin(kz_0)\sinh\left[\kappa(z_c-z_0)\right]\ ,\\
f_{-}(\mu)&= D_1(\mu)e^{-\kappa z_c} - D_2(\mu)e^{\kappa z_c}=-\cos(kz_0)\sinh\left[\kappa(z_c-z_0)\right]+\frac{k}{\kappa}\sin(kz_0)\cosh\left[\kappa(z_c-z_0)\right]\ .
\end{align}
\end{subequations}
\end{widetext}
The remaining overall constant $u_\mu(0)$, and consequently the spectral density~$\rho(\mu)=|u_{\mu}(0)|^2$ can be determined from the normalization condition:
\begin{equation}
\int_{-\infty}^\infty dz\  u_{\mu}(z) u^{*}
_{\mu^\prime}(z)=\delta(\mu-\mu^\prime)\ ,
\end{equation}
which gives
\begin{equation}
\label{rho}
\rho(\mu)=\frac{\mu}{2}\frac{1}{|B_1(\mu)|^2+|B_2(\mu)|^2}\ .
\end{equation}
As we have already stressed in section~\ref{setup},~$\rho(\mu)$ is the quantity relevant to  the effective gravity on the brane and it is therefore useful  to examine its behavior as a function of~$\mu$ in some detail.

In the following, we assume that~$b(z_c-z_0)\gg 1$, that is that the tunneling probability of the light modes through the potential barriers is very small. We suppose also that the constant~$z_1$ is adjusted in such a way that the ratio~$(z_c-z_1)/(z_c-z_0)\ll 1$. 

For~$\mu\ll \mu_1\equiv\min\left(a,b,(z_c-z_1)^{-1}\right)$, we can approximate~$B_1(\mu)$ and~$B_2(\mu)$ by their  power series in~$\mu$:
\begin{eqnarray}
B_1(\mu) &=& a_0 \mu^{-\nu}+a_1\mu^{-\nu+2}+\dots+b_0 \mu^{\nu}+\dots \\
B_2(\mu) &=& c_0\mu^\nu+\dots 
\end{eqnarray}
The condition of  continuity  of the superpotential~$(\ref{cont_w_2})$ forces~$a_0$ to vanish and guarantees that~$c_0$ and $b_0$ are small (of order $\exp[-b(z_c-z_0)]$). 
 As a consequence~we have for $\rho(\mu)$:
\begin{equation}
\label{C_approx}
\rho(\mu)\approx\frac{\sin^2(\nu\pi)}{\pi^2}\frac{1}{\mu (z_c-z_1)}\frac{A(\mu)}{\alpha_0^2+\alpha_1^2A(\mu)+\alpha_2^2 A(\mu)^2}\ ,
\end{equation}
where $A(\mu)$ stands for:
\begin{equation}
A(\mu)=\left[\mu(z_c-z_1)\right]^{2-2\nu}e^{2b(z_c-z_0)}
\end{equation}
and
\begin{equation}
\alpha_0^2 = \frac{2^{3-2\nu}}{\Gamma(\nu)^2}\frac{a^2b^2}{a^2+b^2}\left[b+\frac{c}{z_c-z_1}\right]^{-2}\ , \qquad 
\alpha_1^2 = -\frac{\sin(2\nu\pi)}{2\pi b}\frac{1+bz_0}{z_c-z_1} \ , \qquad
\alpha_2^2 = \frac{1}{4\cos^2(\nu\pi)}\frac{\alpha_1^4}{\alpha_0^2} \ .
\end{equation}
The equality~$(\ref{C_approx})$ shows that for small $\mu$'s the spectral density~$\rho(\mu)$ is  a uniformly decreasing function of~$\mu$, strongly peaked at~$\mu=0$ where it has a singularity of the form $\mu^{(1-2\nu)}$. The bulk of the weight of~$\rho(\mu)$ lies in the region between~$\mu=0$ and  $\mu\sim\mu_1\exp\left(-\frac{b}{1-\nu}(z_c-z_0)\right)$ where~$A(\mu)\sim {\cal O}(1)$. We have therefore in our problem two  characteristic mass scales, $\mu_1=\min\left(a,b,(z_c-z_1)^{-1}\right)$ and~$\mu_2=\mu_1\exp\left(-\frac{b}{1-\nu}(z_c-z_0)\right)$. The assumption of small tunneling probability ensures that these scales are well separated,~$\mu_2\ll\mu_1$. 

\noindent Let us remark that in the limit~$\nu=1/2$ ,
we  recover the results of~\cite{ceh} and $\rho(\mu)$ assumes the Breit-Wigner form:
\begin{equation}
\rho(\mu)\approx\frac{\cal{A}}{\mu^2+\Delta\mu^2}\ ,
\end{equation}
with the resonance width given by
\begin{equation}
\Delta\mu\approx \frac{8}{\left(1+b^2/a^2\right)\left(z_0+1/b\right)}e^{-2b(z_c-z_0)}\ .
\end{equation}

To conclude this chapter, let us discuss briefly the  connection between the spectral  density~$\rho(\mu)$ and  
the scattering matrix of our one-dimensional problem.
The relevant eigenvalue of the $S$-matrix is:
\begin{equation}
\label{s-matrix}
S_+(\mu)=\frac{B_1(\mu)-iB_2(\mu)}{B_1(\mu)+i B_2(\mu)}e^{-2i\left(\frac{\pi}{2}\nu+\frac{\pi}{4}+\mu z_1\right)}=\frac{\Phi(\mu)}{{\Phi}^{*}(\mu)}\ ,
\end{equation}
where the Jost function $\Phi(\mu)$ is defined in such a way that  tends to one in the non-interaction limit and is given by:
\begin{eqnarray}
\Phi(\mu)&=&\sqrt{\frac{2}{\pi\mu}}\,e^{-i\left(\frac{\pi}{2}\nu+\frac{\pi}{4}+\mu z_1\right)}\left[B_1(\mu)-iB_2(\mu)\right]\ . 
\end{eqnarray}
Comparing the expression for the spectral density~$(\ref{rho})$ to~$(\ref{s-matrix})$, we see that $\rho(\mu)$ can be written as
$$
\rho(\mu)=\frac{\pi}{\Phi(\mu)\Phi^{*}(\mu)}\ .
$$
 It is  known from the formal scattering theory~\cite{taylor} that physical properties of a quantum system are closely related with the analytic properties of its scattering matrix. In particular, poles of the $S$-matrix in the complex $\mu$-plane correspond, depending on their location, to bound states, virtual states or resonances  of the system.
In our problem, $S$-matrix has in general  a branch point singularity at $\mu$=0, which it  inherits from the Bessel functions $J_\nu(\mu(z_c-z_1))$ and~$N_\nu(\mu(z_c-z_1))$ entering the expression $(\ref{s-matrix})$. The results of the next section will demonstrate  that this type of singularity can also strongly influence the physics.

\section{Effective gravity}

Let us now turn to the contribution of the tensor modes to the effective gravity on the brane.
The static potential between two sources on the brane, separated by the distance~$r\equiv|\vec{x}-\vec{x}^\prime|$
receives Yukawa-type contributions from  all  the modes and is given by:
\begin{equation}
V_G(r)=\frac{G_5}{4\pi}\int_0^\infty d\mu\ \frac{e^{-\mu r}}{r} |u_{\mu}(0)|^2 =\frac{G_5}{4\pi}\int_0^\infty d\mu\ \frac{e^{-\mu r}}{r} \rho(\mu)\ .
\end{equation}
It is convenient to divide the integral into two parts:
\begin{equation}
V_G(r)=\frac{G_5}{4\pi}\int_0^\infty d\mu\ \frac{e^{-\mu r}}{r}\rho(\mu) =\frac{G_5}{4\pi}\int_0^{\mu_1} d\mu\ \frac{e^{-\mu r}}{r}\rho(\mu)+\frac{G_5}{4\pi}\int_{\mu_1}^\infty d\mu\ \frac{e^{-\mu r}}{r} \rho(\mu) \ .
\end{equation}
For distances~$r\gg r_{1}={\mu}_1^{-1}$ the second integral is negligible 
 and in the first one we can replace~$\rho(\mu)$ by its approximate form~$(\ref{C_approx})$, which gives us:
\begin{equation}
\label{V_approx}
V_G(r)=G_5\frac{\sin^2(\nu\pi)}{4\pi^3 (z_c-z_1)}\int_0^{\mu_1} d\mu\ \frac{e^{-\mu r}}{\mu r}\frac{A(\mu)}{\alpha_0^2+\alpha_1^2A(\mu)+\alpha_2^2 A(\mu)^2}\ .
\end{equation}
This integral is always saturated for~$A(\mu)\sim {\cal O}(1)$, that is for 
$\mu\sim \mu_2\ll \mu_1$. 
We can therefore extend the integration to infinity and evaluate the resulting   integral in two regimes: 
$r_{1}\ll r\ll r_2=\mu_2^{-1}$ and $r\gg r_2$.
In the first regime, that is for distances $r_{1}\ll r\ll r_2$, the exponential in~$(\ref{V_approx})$ can be set to 1 and we obtain:
\begin{equation}
V_G(r)=\frac{G_5}{4 \pi r}\int_0^\infty  \frac{d\mu}{\mu}\ \frac{\sin^2(\nu\pi)}{\pi^2 (z_c-z_1)}\frac{A(\mu)}{\alpha_0^2+\alpha_1^2A(\mu)+\alpha_2^2 A(\mu)^2} =\frac{\left({\cal{C}}^\infty_0\right)^2}{4\pi}\frac{G_5}{r} \ ,
\end{equation}
where ${\cal{C}}^\infty_0$ is defined by~$(\ref{C_infty})$. Therefore, in this (large) interval of distances, light modes of the continuum spectrum reproduce precisely the same~$1/r$ potential we would get  in the presence of a  massless boson  (or, in other words, a normalizable zero mode).  
  
In the second regime, that is for large distances, $r\gg r_2$, only  $A\ll 1$ will give significant contributions to the integral. We can therefore neglect $A(\mu)$ and $A(\mu)^2$ terms in the denominator of the integrand in~$(\ref{V_approx})$, thus obtaining:
\begin{align}
V_G(r)&=\frac{\sin^2(\nu\pi)}{4\pi^3\alpha_0^2}(z_c-z_1)^{1-2\nu}e^{2b(z_c-z_0)}\frac{1}{r}\int_0^\infty d \mu \  e^{-\mu r}\mu^{1-2\nu}\nonumber \\ 
&=\frac{\sin^2(\nu\pi)}{4\pi^3\alpha_0^2}(z_c-z_1)^{1-2\nu}e^{2b(z_c-z_0)}\Gamma(2-2\nu)\frac{G_5}{r^{3-2\nu}}\ .
\end{align}
Hence, we have a modification of the~$1/r$ law for large distances, with the potential of type~$r^{-\beta}$,where~$\beta\in(1,2)$. Thus, the gravity does not in general become five-dimensional, except in an  asymptotically flat geometry ($\nu=1/2$), where the potential is:
\begin{equation}
V_G(r)=\frac{a^2+b^2}{16\pi^2 a^2}e^{2b(z_c-z_0)}\frac{G_5}{r^{2}} \ .
\end{equation}

Following~\cite{grs}, we can also study the  propagation of  gravitational waves generated by a periodic point-like source on the brane, $T(x,z)=T(\vec{x})e^{-i\omega t}\delta(z)$ (we omit the four-dimensional indices). The  field  induced on the brane is given by the convolution of the source with the Green's function~$(\ref{green_f})$
\begin{equation}
G_R(\vec{x}-\vec{x}^\prime;\omega)=\int_{-\infty}^{\infty}d(t-t^\prime) \ G_R(x,z=0;x^\prime,z^\prime=0)\ e^{-i\omega (t-t^\prime)}\ , 
\end{equation}
which after inserting~$(\ref{green_f})$ 
and simplifying becomes
\begin{equation}
\label{G_omega}
G_R(\vec{x}-\vec{x}^\prime;\omega)=\frac{1}{4\pi}\int_0^\infty  d\mu\  \frac{e^{i\omega_\mu r}}{r}\ \rho(\mu) \ ,
\end{equation}
where again $r\equiv|\vec{x}-\vec{x}^\prime|$ and~$\omega_\mu=\sqrt{\omega^2-\mu^2}$ when~$\mu<\omega$ and~$\omega_\mu=i\sqrt{\mu^2-\omega^2}$ when~$\mu>\omega$. Only modes with~$\mu<\omega$ are actually radiated; the other ones exponentially fall off from the source. Being interested in the propagation of waves, we can integrate in~$(\ref{G_omega})$ only up to~$\mu=\omega$. Considering the range of frequencies $r_2^{-1}\ll\omega\ll r_1^{-1}$
 allows us to replace~$\rho(\mu)$ by~$(\ref{C_approx})$ and approximate~$\omega_\mu$ by~$\omega-\frac{\mu^2}{2\omega}$:
\begin{equation}
\label{G_omega_approx}
G_R(\vec{x}-\vec{x}^\prime;\omega)=\frac{\sin^2(\nu\pi)}{4\pi^3}\frac{e^{i\omega r}}{r}\int_0^\infty \ d\mu \ e^{-i(\mu^2/2\omega)r}\frac{A(\mu)}{\alpha_0^2+\alpha_1^2A(\mu)+\alpha_2^2 A(\mu)^2} \ , 
\end{equation}
where we have extended the integration to infinity, using the fact that the integral is saturated for~$\mu\sim \mu_2\ll\omega$. 

For distances $r\gg 2\omega\, r_2/r_1$
 the phase factor in~$(\ref{G_omega_approx})$ varies very slowly and can be set to 1. We then obtain the usual~$1/r$ dependence of wave amplitude on the distance to the source:
\begin{equation}
G_R(\vec{x}-\vec{x}^\prime;\omega)= \frac{\left({\cal{C}}^\infty_0\right)^2}{4\pi}\frac{e^{i\omega r}}{r}\ . 
\end{equation}
For distances $r\gg 2\omega\, r_2/r_1$
 we can neglect $A(\mu)$ and $A(\mu)^2$ terms in the denominator of the integrand in~$(\ref{G_omega_approx})$, obtaining
\begin{eqnarray}
G_R(\vec{x}-\vec{x}^\prime;\omega)&=&\frac{\sin^2(\nu\pi)}{4\pi^3\alpha_0^2}\Gamma(1-\nu)(z_c-z_1)^{1-2\nu}(2\omega)^{1-\nu}e^{2b(z_c-z_0)} \frac{e^{i\omega r+ i \pi(\nu -1)/2}}{r^{2-\nu}} \nonumber\\
&=& (z_c-z_1)^{1-2\nu}\frac{(2\omega)^{1-\nu}}{\Gamma(1-\nu)}\frac{e^{2b(z_c-z_0)}}{4\pi\cos^2(az_0)}\frac{e^{i\omega r+ i \pi(\nu -1)/2}}{r^{2-\nu}} \ ,
\end{eqnarray}
that is, the amplitude is proportional to~$r^{-\sigma}$, where~$\sigma\in(1,\frac{3}{2})$. This behavior indicates  dissipation of the gravitational waves into the extra dimension.

\section{Conclusions}

In this paper we have studied  a toy model for quasilocalization of gravity
on a brane embedded in an asymptotically warped space-time, restricting our attention to the 
tensor sector of  metric perturbations. While, admittedly, our model is only of limited use as a model for gravity, it allows to uncover quite peculiar behavior of the gravitational potential 
(and related quantities) induced by the tensor modes. In the absence of a normalized zero-mode, the effective potential is of the 4D form $1/r$, regardless of
 whether or not the embedding space-time  is asymptotically flat. At large scales, the presence of the extra dimensions becomes significant and the potential gets modified. The nature of
this modification depends on the asymptotic geometry of the space-time and potential does not necessarily assume the  5D form $1/r^2$.

Although we did not present a concrete model which would possess
asymptotically non-flat solutions with quasilocalized gravity, some
features inherent to such models can be derived. By Einstein equations, the
warped geometry~(\ref{metric}) corresponds to the matter energy-momentum
tensor which obeys 
\begin{equation}
\kappa^2\left(\rho + p_z\right) = {\rm e}^{A(z)}\left( \frac{3}{4} A'^2 + \frac{3}{2} A''\right), 
\label{energy-cond}
\end{equation}
where $\rho$ and $p_z$ are the energy density and pressure in the
$z$ direction, respectively. Making use of Eq.~(\ref{w}) at large $z$ we
find that our asymptotically non-flat ansatz requires $\rho + p_z<0$ and
thus violates the weak energy condition. In this respect, our model is
quite similar to (and may share the problems of) that of
Ref.~\cite{grs}. It  remains yet to be seen whether these problems can be
overcome. This is related to an important question, namely the impact of the scalar sector 
of metric perturbations on the effective gravity in a theory where the embedding space-time  is not 
 asymptotically flat and the graviton is  quasi-localized. It is legitimate to suspect 
that it could be as important  as in the model of~Ref.~\cite{grs} and this matter should be 
 examined on a concrete brane setup.

Finally, it is worth pointing out that Eq.~(\ref{schro}) which is the
starting point of our analysis 
is quite universal, in that it describes -- with a suitably adjusted potential --  
 the localization of matter fields as well. Therefore, our conclusions are directly applicable to the case of
matter fields whenever the corresponding potential  can be cast in a 
``supersymmetric'' form~$V(z)=W^2(z)-W^\prime(z)$.
The latter condition is equivalent to
the requirement that the four-dimensional particle is massless (the massive
case was treated in Ref.~\cite{drt}). In the case of matter,
the potential in Eq.~(\ref{schro}) is no longer related to the metric of 
space-time through the relation~(\ref{pot})
 and the problem of violation of the weak energy condition does not need to arise.

\begin{acknowledgments}

The work of P. T. and M. S. is  supported in part by  Swiss National Science Foundation. 
\end{acknowledgments}

\end{document}